\documentclass[preprint]{revtex4}
\usepackage{graphicx}
\usepackage{amssymb}
\newcommand{\be}{\begin{eqnarray}}
\newcommand{\ee}{\end{eqnarray}}
\begin{document}

\title{Twist Three Distribution $f_\perp(x,k^\perp)$ in Light-front Hamiltonian
Approach}
\author{\bf A. Mukherjee and R. Korrapati}
\affiliation{$^a$ Department of Physics,\\
Indian Institute of Technology Bombay, Powai, Mumbai 400076,
India.}
\date{\today}
\begin{abstract}
We calculate the twist three distribution $f_\perp(x,k^\perp)$ contributing
to Cahn effect in unpolarized semi-inclusive deep inelastic scattering.
We use light-front Hamiltonian technique and take the state to be a dressed
quark at one loop in perturbation theory. The 'genuine twist
three' contribution comes from the quark-gluon interaction part in the
operator and is explicitly calculated. $f_\perp(x,k^\perp)$ is compared with
$f_1(x,k^\perp)$.   
\end{abstract}
\maketitle
\section{Introduction}
Transverse momentum dependent parton distributions (TMDs) \cite{tang} 
have gained a lot of
interest recently. In collinear hard scattering processes, for example, 
in deep inelastic scattering (DIS) the large virtuality $Q^2$ of the hard probe
(virtual photon) introduces a longitudinal direction. A plane perpendicular
to that is the transverse plane. Ordinary parton distributions (pdfs)
measured in inclusive processes like DIS do not give any information on the
transverse momentum distributions of quarks and gluons. TMDs can be measured
in processes when sufficient transverse momentum is measured in the final
state; for example in semi-inclusive deep inelastic scattering (SIDIS) where
a hadron with transverse momentum $P_h^\perp$ is measured or in Drell-Yan
process where the transverse momentum of the virtual photon is measured.
Factorization for some processes involving TMDs have been proven at twist
two at one loop order and argued to hold for all orders \cite{fac,fac1}. The 
TMDs involve an operator structure which is bilocal both in the light cone 
as well as
in the transverse direction; and  a path ordered exponential of the line
integral of the gauge field (gauge link) is necessary for color gauge
invariance. In light front gauge,  the link in the light cone direction
becomes unity, but contribution comes from the part of the link at
light-cone infinity involving the transverse component of the gauge field.
It has been found that this is process dependent in general \cite{pijl} 
and may
contribute at leading order in $1/Q$ \cite{final}. However for fragmentation 
process the shape of the Wilson line has no effect on observables
\cite{frag}. In SIDIS and Drell-Yan the TMDs are simply connected by 
a reversal of sign \cite{sign}. In more
complicated processes like hadron production in hadron-hadron collisions,
although  the standard universality of TMDs does not hold, predictivity is
not lost \cite{hh}. However, very recently in \cite{ted} it has been shown 
that  such 'generalized' factorization does not hold for all 
hadroproduction processes. 

There are $32$ quark TMDs including twist two, three and four. The higher
twist or subleading in $1/Q$  TMDs contain one or more 'bad' light cone
component of the quark field, and the operator involves quark-gluon
interaction term. The subleading twist TMDs are important as they contribute
in several single spin as well as azimuthal asymmetries in the kinematical
range of present experiments. Experimental data on several of these
asymmetries are now available \cite{exp}. Interpretation of these 
subleading twist asymmetries are more challenging as they involve 
several higher twist
distribution and fragmentation functions. Here model calculations of these
functions play an important role. Twist three TMDs are related to twist two
TMDs and 'genuine twist three part' through equation of motion relations
\cite{eqm}. In certain models some other relations between the 
TMDs exist based on Lorentz invariance. The Lorentz invariance 
relations do not hold in QCD due to the presence of the gauge link
\cite{lorentz} whereas the equation of motion relations still hold \cite{eqm}. 

Among the twist three TMDs there are few model calculations of
$f_\perp(x,k^\perp)$. This is time-reversal even and  plays an important 
role in $cos ~ \phi_h$ asymmetry in unpolarized SIDIS, the so-called 
Cahn effect \cite{cahn}. The unpolarized  SIDIS
cross section depends on the azimuthal angle $\phi_h$ between the lepton
plane and the hadron production plane, and on the transverse momentum of the
detected hadron. The $\phi_h$ dependence of the SIDIS cross section has been
experimentally detected by the EMC collaboration \cite{emc}. If one neglects the
explicit quark gluon interaction terms in the distribution and fragmentation
functions then this $cos~\phi_h$ dependence of the cross section is given in
terms of the unpolarized distribution and fragmentation functions at $1/Q$
level. This effect has been investigated in a parton model approach by 
introducing a                           
phenomenologically motivated intrinsic $k^\perp$ dependence \cite{ansel}. 
$f_\perp(x,k^\perp)$ has been calculated in a simple spectator model in 
\cite{spec}. 
A bag model result of $f_\perp(x,k^\perp)$ has been given in \cite{harut}.    

In this work we calculate $f_\perp(x,k^\perp)$ in light front Hamiltonian
approach. Instead of using the Feynman diagrams, we expand the state
in Fock space in terms of multi-parton light-front wavefunctions. 
The partons are on-mass shell interacting objects having non-vanishing transverse
momenta and thus they can be called field theoretic partons. The advantage is that
these wave functions are Lorentz boost invariant \cite{LFWF}, so we can truncate the
Fock space expansion to a few particle sector in a boost invariant way.
We take the state to be a dressed quark at one loop in QCD. 
The two particle light-front wave functions (LFWFs) can be calculated   
analytically for a quark at one loop using the light-front Hamiltonian. 
 Using the constraint
equation in light-front gauge, the bad component of the fermion field, 
$\psi^{(-)}$, is eliminated. The operator has a mass dependent part, a $k^\perp$
dependent part and a quark-gluon interaction part. The distribution can   
be expressed in terms of overlaps of LFWFs. In addition to diagonal overlaps
there are particle number changing off-diagonal overlaps. Twist three
distributions $g_T(x), e(x)$ and $h_L(x)$ have been investigated in this
approach before \cite{gt,ex,hl}. In the next section, we present details of
the calculation. We end with discussions of the result.  

\section{Twist three distribution $f_\perp(x,k^\perp)$}
The transverse momentum dependent distribution $f_\perp(x,k^\perp)$is 
defined as 
\be
{k^i\over P^+} f_\perp(x,k^\perp)= \int
{dy^- d^2 y^\perp \over  4{(2 \pi)}^3} e^{{i\over 2} P^+ y^-x}
 e^{-i k^\perp \cdot
y^\perp}  {\langle P \mid {\bar{\psi}}(0) U (0,y) \gamma^i 
\psi(y^-,y^\perp) \mid P \rangle \mid}_{y^+=0}.
\label{fperpk}
\ee
$U(0,y)$ is the path ordered exponential (link) required for color gauge
invariance. For transverse momentum dependent distributions, the bilocality
in the operator is both in the longitudinal as well as in the transverse
direction. In the light cone gauge, $A^+=0$, the gauge link in the 
longitudinal or light-cone direction becomes
unity,  but contribution will come from the transverse gauge link at light
cone infinity which can not be set to unity in this gauge. It has been found
recently that this part of the gauge link gives important contribution even
at twist two level, in particular in the case of time-reversal odd
observables \cite{pijl}. However, in  the following, we neglect the 
contribution from the transverse gauge link. 

We take $i=1$. We have, using the light-front projection operators $
\Lambda^\pm={1\over 2} \gamma^0 \gamma^\pm$;
\be
{\bar{\psi}}(0) \gamma^i \psi(y^-,y^\perp)=\psi^{(-) \dagger}(0)
\alpha^i \psi^{(+)}(y^-,y^\perp)
+\psi^{(+) \dagger}(0)\alpha^i \psi^{(-)}(y^-,y^\perp).
\label{current}
\ee
In light-front gauge, $A^+=0$, the 'bad' component, $\psi^{(-)}$ 
is constrained,
and the equation of constraint is given by \cite{two}
\be
\psi^{(-)}(y)={1\over i\partial^+} (i \alpha^\perp \cdot \partial^\perp +
g \alpha^\perp \cdot A^\perp + \beta m) \psi^{(+)}(y);
\ee
where the operator ${1\over \partial^+}$ is defined as \cite{two}
\be
{1\over \partial^+} f(x^-)={1\over 4} \int_{-\infty}^{\infty} dy^-
\epsilon(x^--y^-) f(y^-).
\label{pp}
\ee
The antisymmetric step function is given by
\be
\epsilon(x^-)=-{i \over \pi} {\cal{P}} \int {d \omega\over \omega} e^{{i
\over
2} \omega x^-};
\ee
${\cal{P}}$ denotes the principal value. 
Using the equation of constraint the field $\psi^{(-)}$ can be removed. 
The operator has three parts :
\be
O_{k^\perp}= \psi^{(+) \dagger}(0) \Big [ ( \alpha^\perp \cdot
\stackrel{\leftarrow}{
\partial^\perp}) (\stackrel{\leftarrow} {1\over \partial^+}) \alpha^1+
\alpha^1 (\stackrel{\rightarrow}{1\over \partial^+}) (\alpha^\perp \cdot 
\stackrel{\rightarrow}{\partial^\perp}) \Big ] \psi^{(+)}(y);
\label{Ok}
\ee
\be
O_g= g \psi^{(+) \dagger}(0) \Big [ ( \alpha^\perp \cdot A^\perp )
 (\stackrel{\leftarrow} {1\over -i \partial^+}) \alpha^1+
\alpha^1 (\stackrel{\rightarrow}{1\over i \partial^+}) (\alpha^\perp \cdot 
A^\perp ) \Big ] \psi^{(+)}(y);
\label{Og}
\ee 
\be 
O_m= m \psi^{(+) \dagger}(0) \gamma^1 \Big [ (\stackrel{\leftarrow} 
{1\over -i \partial^+}) - (\stackrel{\rightarrow}{1\over i \partial^+})
\Big ] \psi^{(+)}(y).
\label{Om}
\ee
For the dynamical field
$\psi^{(+)}$ we use two component formalism \cite{two}
$\mid P,\sigma \rangle$ is a proton state of momentum $P$ and helicity
$\sigma$. The
state can be expanded in Fock space in terms of multi-parton LFWFs. Instead
of the proton we take the state to be a dressed quark. Fock space expansion
of such a state can be written as : 
\begin{eqnarray}
\mid P, \sigma \rangle && = \phi_1 b^\dagger(P,\sigma) \mid 0 \rangle
\nonumber \\  
&& + \sum_{\sigma_1,\lambda_2} \int
{dk_1^+ d^2k_1^\perp \over \sqrt{2 (2 \pi)^3 k_1^+}}
\int 
{dk_2^+ d^2k_2^\perp \over \sqrt{2 (2 \pi)^3 k_2^+}}
\sqrt{2 (2 \pi)^3 P^+} \delta^3(P-k_1-k_2) \nonumber \\
&& ~~~~~\phi_2(P,\sigma \mid k_1, \sigma_1; k_2 , \lambda_2) b^\dagger(k_1,
\sigma_1) a^\dagger(k_2, \lambda_2) \mid 0 \rangle.
\label{eq2}    
\end{eqnarray} 
Here $a^\dagger$ and $b^\dagger$ are bare gluon and quark
creation operators respectively and $\phi_1$ and $\phi_2$ are the
multiparton wave functions. 
We introduce Jacobi momenta $x_i$,~${q_i}^\perp$ such that 
$\sum_i x_i=1$ and $\sum_i {q_i}^\perp=0$.  They are defined as
\be
x_i={k_i^+\over P^+}, ~~~~~~q_i^\perp=k_i^\perp-x_i P^\perp.
\ee
Also, we introduce the wave functions,
\be
\psi_1=\phi_1, ~~~~~~~~~~~\psi_2(x_i,q_i^\perp)= {\sqrt {P^+}} \phi_2
(k_i^+,{k_i}^\perp);
\ee
which are independent of the total transverse momentum $P^\perp$ of the
state and are boost invariant. The two particle wave function depends 
on the helicities of the quark and gluon. Using the eigenvalue equation 
for the light-cone Hamiltonian, this  can be written as \cite{rajen},
\be
\psi^\sigma_{2\sigma_1,\lambda}(x,q^\perp)&=& {x(1-x)\over
(q^\perp)^2+m^2 (1-x)^2}
{1\over {\sqrt {(1-x)}}} {g\over
{\sqrt {2(2\pi)^3}}} T^a \chi^\dagger_{\sigma_1}\Big [- 2 {q^\perp\over
{1-x}}-{{\tilde \sigma^\perp}\cdot q^\perp\over x} {\tilde \sigma^\perp}
\nonumber\\&&~~~~~~~~~~~~~~~~~~
+i m{\tilde \sigma}^\perp {(1-x)\over x}\Big ]\chi_\sigma
\epsilon^{\perp *}_\lambda \psi_1.
\label{psi2}
\ee
$m$ is the bare mass of the quark, $\tilde \sigma_1=\sigma_2$, $\tilde
\sigma_2=-\sigma_1$. 
$\psi_1$ actually gives the normalization of the state \cite{rajen}:
\be
{\mid \psi_1 \mid}^2=1-{\alpha_s\over {2 \pi}} C_f~log
{Q^2\over \mu^2}
\int_\epsilon^{1-\epsilon} dx {{1+x^2}\over {1-x}},
\label{norm}
\ee
to order $\alpha_s$.  Here $\epsilon$ is a small cutoff on $x$. We have
taken the cutoff on the transverse momenta to be $Q$, 
the large scale of the process. 
In the above expression, we have neglected subleading finite
pieces. $\mu$ is a small scale such that $(q^\perp)^2 > \mu^2 >> m^2$.

Contribution from $O_m$ is  zero. $O_{k^\perp} $ 
has contribution from single particle sector
as well as two particle sector of the state and $O_g$ will get 
contribution from an overlap of a single particle and a two-particle 
light-front wave functions. The contributions from $O_{k_\perp}$ are : 
\be
\int
{dy^- d^2 y^\perp \over  4{(2 \pi)}^3}&& e^{{i\over 2} P^+ y^-x} 
e^{-i k^\perp \cdot
y^\perp}  \langle P\mid O_{k^\perp} \mid P \rangle =
\delta(1-x){P^1\over P^+} \delta^2(k^\perp-P^\perp) {\mid \psi_1 \mid}^2
 \nonumber\\&&
+ \int d^2 q^\perp {q^1+x P^1\over x P^+} {\mid
\psi_{2,s_1,\lambda}^s(x,q^\perp)\mid}^2 \delta^2(k^\perp-q^\perp-x P^\perp) 
\nonumber\\&& 
={P^1\over P^+} \delta(1-x) \delta^2(k^\perp-P^\perp){\mid \psi_1 \mid}^2
\nonumber\\&&+{\alpha_s\over 2
\pi^2} C_f \int d^2 q^\perp \delta^2 (k^\perp-q^\perp-x P^\perp)
{[{(q^\perp)}^2 ({1+x^2\over 1-x}) +m^2 (1-x)^3]\over {[{(q^\perp)}^2 +m^2
(1-x)^2]}^2 } {(q^1+x P^1)\over xP^+}
\label{nonflip}
\ee
Here we have summed over the helicity of the state. 
The Fock space expansion of the interaction part of the operator can be
written as
\be
O_g^{(1)}&=& g \sum_{spins} \int (dk_1) \int (dk_2) \int [dk_3]~
\chi^\dagger_{\lambda_1}~ \Big ( \epsilon^1_{\lambda_3} -i \sigma_3
\epsilon^2_{\lambda_3} \Big ) \chi_{\lambda_2}~ e^{-{i\over 2} k_2^+
y^-+i k_2^\perp \cdot y^\perp}\nonumber\\&&
\Big [ {1\over k_1^+-k_3^+} b^\dagger_{\lambda_1}(k_1) b_{\lambda_2}(k_2)
 a_{\lambda_3}(k_3) + {1\over k_1^++k_3^+} b^\dagger_{\lambda_1}(k_1)
 b_{\lambda_2}(k_2) a^\dagger_{\lambda_3}(k_3) \Big ]
\ee   
\be
O_g^{(2)}&=& g \sum_{spins} \int (dk_1) \int (dk_2) \int [dk_3]~
\chi^\dagger_{\lambda_1}~ \Big ( \epsilon^{*1}_{\lambda_3} +i \sigma_3
\epsilon^{*2}_{\lambda_3} \Big ) \chi_{\lambda_2}~\nonumber\\&&
 e^{-{i\over 2} k_2^+ y^-+i k_2^\perp \cdot y^\perp}
\Big [ {1\over k_2^++k_3^+} b^\dagger_{\lambda_1}(k_1) b_{\lambda_2}(k_2)
 a_{\lambda_3}(k_3)e^{-{i\over 2} k_3^+y^-+i k_3^\perp \cdot y^\perp}
 \nonumber\\&&~~~~~~~~~~~
+ {1\over k_2^+-k_3^+} b^\dagger_{\lambda_1}(k_1)
 b_{\lambda_2}(k_2) a^\dagger_{\lambda_3}(k_3)e^{{i\over 2} k_3^+ y^--
i k_3^\perp \cdot y^\perp} \Big ].
\ee
Here we have used the notations $(dk)={dk^+d^2k^\perp\over 2 {(2 \pi)}^3
\sqrt{k^+}}$ and $[dk]={dk^+d^2k^\perp\over 2 {(2 \pi)}^3 k^+}$.
As we stated above, the interaction part of the operator gives $\psi_1^*
\psi_2$ and  $\psi_2^* \psi_1$ type terms. 
Contribution from $O_g$ is given by :
\be
\int
{dy^- d^2 y^\perp \over  4{(2 \pi)}^3}&& e^{{i\over 2} P^+ y^-x} 
e^{-i k^\perp \cdot y^\perp} 
 \langle P \mid O_g \mid P \rangle = 
-{\alpha_s\over 2 \pi^2} C_f \int d^2 q^\perp {q^1\over P^+} {1\over 
{[(q^\perp)}^2+m^2 (1-x)^2]} \nonumber\\&&
{1\over x (1-x)} \delta^2 (k^\perp-q^\perp-x P^\perp).
\ee      
Here we have used explicit form of $\psi_{2,s_1,\lambda}^s(x,q^\perp) $. 

In the frame where $P^\perp=0$, one has
\be
x f_\perp(x,k^\perp) = 
{\alpha_s\over 2 \pi^2}\Big ({{(k^\perp)}^2 {(1+x^2)/(1-x)} +m^2 (1-x)^3\over 
{[{(k^\perp)}^2 +m^2 (1-x)^2]}^2 }
-{1\over {(1-x)[{(k^\perp)}^2 +m^2 (1-x)^2]}} \Big ) 
\nonumber\\ 
={\alpha_s\over 2 \pi^2}  
{[{(k^\perp)}^2 x^2  +m^2 (1-x)^2 (x-2) x]\over 
{(1-x) [{(k^\perp)}^2 +m^2 (1-x)^2]}^2 }. 
\ee 
The twist two unpolarized distribution $f_1(x,k^\perp)$ can be calculated
using the definition 

\be
 f_1(x,k^\perp)= \int
{dy^- d^2 y^\perp \over  4{(2 \pi)}^3} e^{{i\over 2} P^+ y^-x}
 e^{-i k^\perp \cdot
y^\perp}  {\langle P\mid {\bar{\psi}}(0) U (0,y) \gamma^+ 
\psi(y^-,y^\perp) \mid P \rangle \mid}_{y^+=0}.
\label{f1}
\ee
The operator neglecting the gauge link is of the form $2 \psi^{(+)\dagger}
(0 ) \psi^{(+)}(y^-, y^\perp)$. 
For a dressed quark state, one gets
\be
f_1(x,k^\perp)&=& \delta(1-x) \delta^2 (k^\perp-P^\perp) {\mid \psi_1\mid
}^2\nonumber\\&&+{\alpha_s\over 2
\pi^2} C_f \int d^2 q^\perp \delta^2 (k^\perp-q^\perp-x P^\perp)
{[{(q^\perp)}^2 ({1+x^2\over 1-x}) +m^2 (1-x)^3]\over {[{(q^\perp)}^2 +m^2
(1-x)^2]}^2 }.
\ee
In the frame $P^\perp=0$ one gets
\be
f_1(x,k^\perp)&=& {\alpha_s\over 2 \pi^2} C_f 
{[{(k^\perp)}^2 ({1+x^2\over 1-x}) +m^2 (1-x)^3]\over {[{(k^\perp)}^2 +m^2
(1-x)^2]}^2 };
\ee
neglecting the single particle contribution at $x=1$ and $k^\perp =0$.
The above result agrees with \cite{quarktarget}. Note that in order to get
the correct behaviour at $x=1$ one has to include the single particle
contribution and the normalization of the state Eq. (\ref{norm}). 
Comparing we see the equation of motion relation 
\be
x f_\perp=x \tilde f_\perp + f_1 
\ee
is satisfied, with $\tilde f_\perp$ is the genuine twist three 
quark-gluon interaction part which in our calculation, comes from $O_g$. 

In the TMDs we did not use the large $k^\perp$ approximation.
However in the limit of large $k^\perp$, the twist three distribution has
$1\over {(k^\perp)}^2$ behaviour as shown in \cite{large}.  

\begin{figure}[!htp]
\begin{minipage}[c]{0.9\textwidth}
\tiny{(a)}\includegraphics[width=6.5cm,height=5cm,clip]{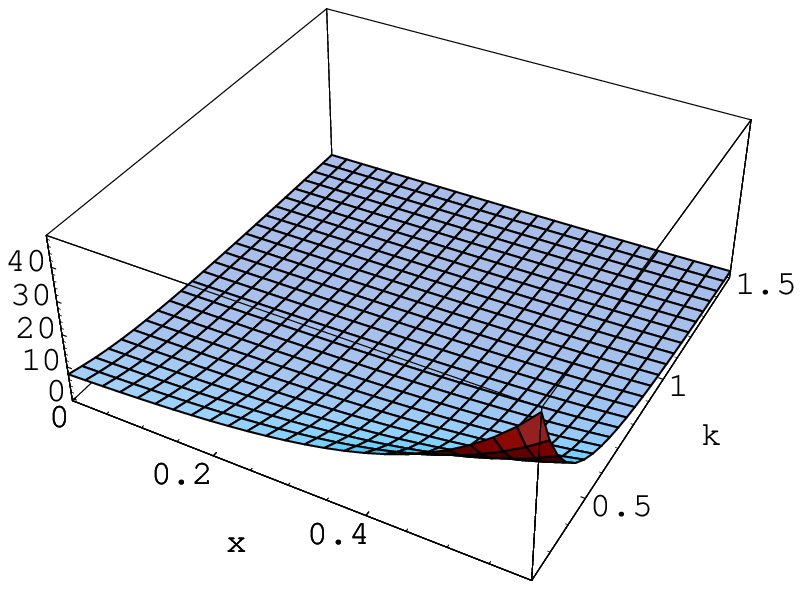}
\hspace{0.1cm}%
\tiny{(b)}\includegraphics[width=6.5cm,height=5cm,clip]{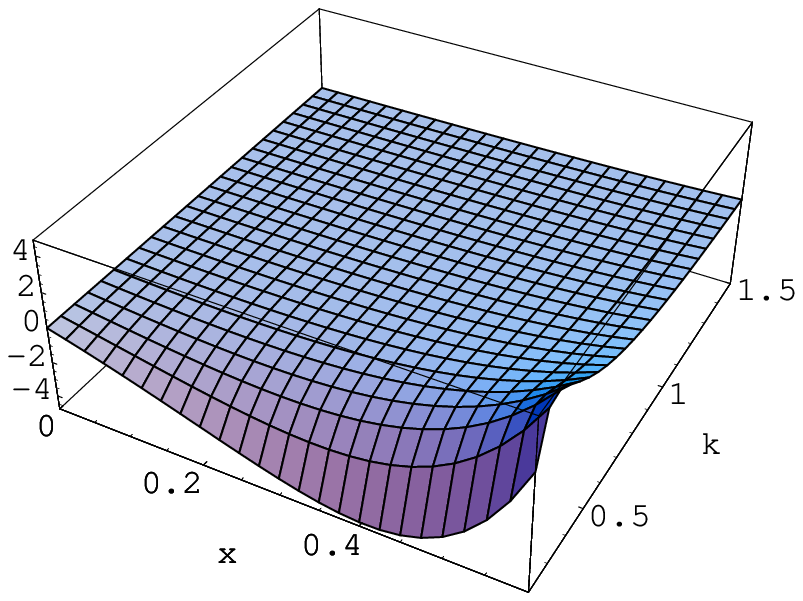}
\end{minipage}
\caption{\label{plot}(Color online)  Plots of (a) $f_1(x,k^\perp)$ 
and (b) $x f_\perp(x,k^\perp)$ vs $x$ and $k={\mid k^\perp \mid}$. 
We have taken $m=0.3$ GeV. $k^\perp$ is in GeV.} 
\end{figure}

In Fig. 1 we have plotted $f_1(x,k^\perp)$ and $xf_\perp(x,k^\perp)$
as functions of $x$ and $k^\perp$. Substantial difference is observed in
relatively lower $k^\perp$ region, in fact $x f_\perp$ also
becomes negative. This is due to the
quark-gluon interaction contribution to the twist three distribution, and
unlike the bag model \cite{harut}.   
We took $m=0.3$ GeV. We have divided both plots by ${\alpha_s\over 2
\pi^2}$. One has to be careful not to compare the numerical results of the
dressed quark calculations with experimental data. However, the qualitative
behaviour is interesting as unlike phenomenological models, the genuine
twist three part comes from explicit calculation of the quark gluon
interaction term.

In the integrated distribution, there is an integration over
$k^\perp$. The
operator is bilocal only in minus direction. As a result, the gauge link is
only in the light-cone direction and becomes  unity in the light cone gauge.
The operator can still be separated into three parts, $O_m$, $O_{k^\perp}$
and $O_g$ using the equation of constraint for $\psi^{(-)}$. 
$O_m$, as before gives zero contribution. Contribution of $O_g$ is zero 
after $k^\perp$ integration, due to rotational symmetry. The entire 
contribution comes from $O_{k^\perp}$ :
\be
\int {dy^-\over 8 \pi} e^{i P^+ x y^-\over 2} 
\langle P \mid O_{k^\perp} \mid P\rangle =  
{P^1\over  P^+} \Big [\delta(1-x)+{\alpha_s \over 2 \pi} log{Q^2 \over \mu^2}
C_f \int dx {1+x^2\over 1-x}_+ \Big ].
\label{int}
\ee
{\it rhs} is the twist two
unpolarized distribution function $f_1(x, Q^2)$. This is expected as when 
integrated distributions are concerned, the transverse component of the
bilocal current given by Eq. (\ref{current}) with bilocality only in the
minus direction, has the same parton interpretation as the plus component. 
Note that we get nonzero result only when $P^\perp$ is nonzero \cite{rajen}. 
\section{Discussion}
In this paper, we calculate the twist three distribution
$f_\perp(x,k^\perp)$ in light-front Hamiltonian approach. This distribution
is known to play an important role in the observed Cahn effect in
unpolarized SIDIS. Instead of a proton state we take the state to be a
dressed quark at one loop in QCD. The advantage is that the higher Fock
space component (two particle) LFWF can be calculated analytically. These
play an important role in the higher twist distributions. The partons, that
is, the quarks and gluons have non-zero transverse momenta and they
interact. The transverse momentum dependence of the two-particle LFWF 
is obtained by solving the eigenvalue equation of the light-front
Hamiltonian. At $O(\alpha_s)$
this calculation is exact. However, we neglect the contribution from the gauge
link at light-cone infinity. The operator has three parts, an intrinsic
transverse momentum dependent term, a mass term and a 'genuine twist three'
quark-gluon interaction term. Contribution from each of these terms are
calculated using overlaps of LFWFs. The equation of motion relation
connecting $f_\perp(x,k^\perp)$ to the twist two unpolarized distribution
$f_1(x,k^\perp)$ and a quark-gluon interaction part is shown to hold.
$x f_\perp(x,k^\perp)$ differs substantially in qualitative behaviour from
$f_1(x,k^\perp)$ in low $k^\perp$ region.  
The last part vanishes when integrated ove $k^\perp$ and one gets the same
information as in $f_1(x,Q^2)$.       
\section{acknowledgment}
This work is
supported by BRNS grant Sanction No. 2007/37/60/BRNS/2913 dated 31.3.08,
Govt. of India.

\end{document}